# Stacking-Dependent Electronic Property of Trilayer Graphene Epitaxially Grown on Ru(0001)

Yande Que, Wende Xiao,[a] Hui Chen, Dongfei Wang, Shixuan Du and Hong-Jun Gao [a]

*Institute of Physics and University of Chinese Academy of Sciences, Chinese Academy of Sciences, Beijing, 100190, China*

**Abstract**

The growth, atomic structure and electronic property of trilayer graphene (TLG) on Ru(0001) were studied by low temperature scanning tunneling microscopy and spectroscopy (LT-STM/STS) in combined with tight-binding approximation (TBA) calculations. TLG on Ru(0001) shows a flat surface with a hexagonal lattice due to the screening effect of the bottom two layers and the AB-stacking in the top two layers. The coexistence of AA- and AB-stacking in the bottom two layers leads to three different stacking orders of TLG, namely, ABA-, ABC- and ABB-stacking. STS measurements combined with TBA calculations reveal that the density of states of TLG with ABC- and ABB-stacking are characterized by one and two sharp peaks near to the Fermi level, respectively, in contrast to the V-shaped feature of TLG with ABA-stacking. Our work demonstrates that TLG on Ru(0001) might be an ideal platform for exploring stacking-dependent electronic properties of graphene.

a) To whom correspondence should be addressed. Tel.: +86-10-82648048, Fax: +86-10-62556598, Electronic mail: wdxiao@iphy.ac.cn and hjgao@iphy.ac.cn

Within a decade after the discovery of graphene flakes by mechanical exfoliation,[1] graphene based structures have drawn numerous attentions due to their unique electronic properties and potential applications.[2-4] In multilayer graphene, the stacking configuration provides an important degree of freedom for tuning its electronic properties. For instance, Bernal (AB-stacked) bilayer graphene (BLG) exhibits a quadratic energy–momentum dispersion due to interlayer interaction,[5] while twisted BLG may show von Hove singularity and superlattice Dirac cone in its density of states (DOS) around the Fermi level ($E_F$) due to the modulation of moiré patterns arising from in-plane rotation between the two graphene layers.[6,7] Based on the Bernal stacking of adjacent layers, two natural stable allotropes of trilayer graphene (TLG) can be constructed, one with the top layer directly lying above the bottom layer (Bernal or ABA-stacked), and the other with one sublattice of the top layer lying above the center of the hexagon of the bottom layer (rhombohedral or ABC-stacked). In recent years, both ABA- and ABC-stacked TLG were obtained via mechanical exfoliation of graphite[8-11] and epitaxial growth on SiC.[12] It has been shown that ABA-stacked TLG is semi-metallic with a tunable band overlap,[8,13] whereas ABC-stacked TLG exhibits flat electronic spectrum in low energy,[14,15] which is predicted to result in many exciting properties, e.g. spontaneous band gap opening,[16] superconductivity,[17] and ferromagnetism.[18] Although large-area high-quality single-layer (SLG)[19-23] and BLG[24,25] have been epitaxially grown on various transition metals and their electronic structures have been investigated by state-of-the-art scanning tunneling microscopy/spectroscopy (STM/STS) at the atomic level, the growth of TLG on transition metals and their physical properties have been rarely addressed.[26]

Previously, we reported on the growth and structural properties of large-area BLG on Ru(0001) surface by low temperature (LT) STM/STS.[25] We found that the lattice of the bottom layer of BLG is stretched by 1.2%, while strain is absent from the top layer. The lattice mismatch between the two layers leads to the formation of a moiré pattern with a periodicity of ~21.5 nm and a gradual change from AB- to AA- stacking in each unit cell of this moiré superstructure in BLG. In this work, we demonstrate by LT-STM/STS that TLG grown on Ru(0001) exhibits three different stacking orders, namely, ABA-, ABC- and ABB-stacking, due to the AB-stacking in the top two layers and the coexistence of AA- and AB-stacking in the bottom two layers. STS measurements combined with tight-binding approximation (TBA) calculations disclose that different stacking order in TLG results in significant difference in electronic structure near to the Fermi level.

Our experiments were carried out in an ultrahigh vacuum (base pressure of $1 \times 10^{-10}$ mbar) LT-STM system (Unisoku), equipped with standard surface preparation facilities including an ion sputtering gun and electron-beam heater for surface cleaning. The Ru(0001) substrate was cleaned by repeated cycles of ion sputtering using $Ar^+$ with an energy of 1.5 keV, annealing at 1400 K and oxygen exposure at 1200 K ($5 \times 10^{-7}$ mbar, 5 min). Prior to the growth of graphene, the surface cleaning of the Ru(0001) substrate was checked by low energy electron diffraction and STM. TLG islands surrounding with BLG was grown by exposing the clean Ru(0001) substrate to 150 L ethylene ($2 \times 10^{-6}$ mbar, 100 s) at 1400 K, followed with cooling down to room temperature with a rate of ~60 K/min. It is noteworthy that an elevated growth temperature of 1400 K increases carbon solubility in bulk Ru, favoring the growth of TLG islands. This process is very similar to that for growing BLG,[25] except with double dosage of ethylene. STM images were acquired in the constant-current mode, and all given voltages refer to the sample. Differential conductance (dI/dV) spectra were collected by using a lock-in technique with a 5 $mV_{rms}$ sinusoidal modulation at a frequency of 973 Hz. All STM/STS experiments were performed with electrochemically etched tungsten tips at 4.2 K, which were calibrated against the surface state of the Au(111) surface and the V-shaped DOS of BLG on Ru(0001) before spectroscopic measurements.

Figure 1(a) shows a large-scale STM image of the as-grown graphene on Ru(0001). Two kinds of areas with different morphology can be easily distinguished. The corrugated areas exhibit a variation of ~ 1 Å in apparent height and an ordered moiré pattern with a periodicity of ~ 3 nm, akin to that of SLG or BLG grown on Ru(0001).[19,20,25] Figure 1(b) illustrates a typical atomic-resolution STM image acquired on the corrugated areas. Both hexagonal and honeycomb lattices are resolved in the atop regions of the moiré pattern. In previous work, we have shown that, for SLG grown on Ru(0001) surface, only honeycomb lattice can be resolved in the atop regions of the moiré pattern, due to the preservation of AB-symmetry in these regions. Meanwhile, for BLG grown on Ru(0001) surface, both hexagonal and honeycomb lattices can be seen in the atop regions of the moiré pattern, as the stacking of the two graphene layers is continuously varied from the AA to AB fashion and vice versa, due to lattice mismatch between the stretched bottom layer and the free-standing top layer. Thus, we assign these corrugated areas of the as-grown graphene to BLG rather than SLG.

Figure 1(c) shows an atomic-resolution STM image acquired on the flat areas of the as-grown graphene. A hexagonal lattice is clearly seen, corresponding to one of the two sublattices of graphene. This behavior indicates that the flat areas are multilayer graphene with thickness $\geqq$ 3 layers and the top two layers are AB-stacked, so that the AB-symmetry of the top layer graphene is broken. The graphene lattice is continuous between the top layer of the corrugated BLG and the flat area, as shown in Fig. 1(d). Line profile analysis (not shown) reveals a height difference of 1.1 Å between the flat and corrugated areas, identical to that between the interlayer distance in graphite (3.3 Å) and the height of atomic step of Ru(0001) substrate (2.2 Å).[27] Thus, we attribute the flat areas to TLG grown on the lower terrace of an atomic step of Ru(0001) substrate. The top two layers of TLG cross the step and seamlessly connect with BLG grown on the upper terrace. Figure 1(e) presents a structural model of such a TLG island connected with BLG on Ru(0001). We note that no significant corrugation arising from the moiré pattern can be observed on TLG. Previous experiments and calculations reveal that SLG grown on Ru(0001) surface is geometrically corrugated and electronically n-doped, due to the inhomogeneous Ru-C coupling,[28] while the top layer of BLG is free-standing with slight corrugation.[29] Therefore, we attribute the flattening of the top layer of TLG to the screening effect of the bottom two layers in TLG, as it has been evidenced that graphene can serve as buffer layer.[30-32] We have collected STM images from various area of the as-grown graphene. Statistical analysis reveals that the percentage of TLG is ~ 5% and the average lateral size of the TLG islands is a few tens nanometer.

For TLG grown on Ru(0001), its bottom two layers exhibit similar structure to that of BLG.[25,29] Meanwhile, the top two layers are decoupled from the Ru substrate by the bottom layer and exhibit a free-standing feature. Therefore, the top two layers are AB-stacked, similar to that of isolated Bernal BLG. Our previous work reveals that for BLG on Ru the bottom layer is stretched by ~1.2%, due to its strong interaction with Ru substrate,[25] while the top layer is nearly decoupled from the substrate by the bottom layer and exhibit free-standing features. The lattice mismatch between the two layers leads to the formation of an additional moiré superstructure with large periodicity of ~21.5 nm and a gradual change from AB- to AA- stacking in each unit cell of this moiré superstructure.[25] Based on the structure of BLG, TLG can be constructed by covering BLG with an additional free-standing graphene layer. This additional layer must be AB-stacked with respect to the top layer of BLG. Thus, the AB-stacked BLG becomes ABA- or ABC-stacked TLG, while the AA-stacked BLG becomes ABB-stacked TLG. Therefore, three types of stacking orders, namely, ABA-, ABC- and ABB-stacking, coexist in TLG grown on Ru(0001), as schematically shown in Fig. 2. The regions between the ABA-, ABC- and ABB-stacked regions exhibit intermediate (incommensurate) stacking fashions. However, as different regions of TLG grown on Ru(0001) surface exhibit a similar hexagonal lattice, the different stacking order cannot be distinguished from STM images alone.

To identify the stacking order and unveil the electronic structure of TLG on Ru(0001), we measured d$I$/d$V$ spectra on various positions of the as-grown TLG. Three types of representative d$I$/d$V$ spectra collected from different positions can be identified, as displayed in Fig. 3. The d$I$/d$V$ spectrum shown in Fig. 3(a) exhibits a V-shaped feature, similar to that of graphite with Bernal stacking.[33] This behavior indicates an ABA-stacking order of this region. In contrast, the d$I$/d$V$ spectrum shown in Fig. 3(b) exhibits a pronounced peak around $E_F$, while the one illustrated in Fig. 3(c) shows two sharp peaks at -29 mV and 29 mV with a local minimum at $E_F$. The significant difference in electronic structures of these two regions suggests that they are not ABA-stacked TLG.

As mentioned above, TLG grown on Ru(0001) surface exhibits a mixture of ABA-, ABC- and ABB-stacking. To give a clear link between the stacking order and the electronic structure in TLG grown on Ru(0001) surface, we carried out TBA calculations for free-standing TLG with ABA-, ABC- and ABB-stacking orders. For simplicity, we only took the nearest-neighbor (NN) hopping of π-electrons into account. The Hamiltonian for these types of TLG can be written as

$$H = \sum_{i=1}^{3} H_i + H_{12} + H_{23} \qquad (1)$$

$$H_i = -t \sum_{k,\sigma} \left( s a^+_{i,k,\sigma} b_{i,k,\sigma} + h.c. \right) \quad (2)$$

$$s = 1 + e^{i\vec{k} \cdot \vec{a}_1} + e^{i\vec{k} \cdot \vec{a}_2} \quad (3)$$

where $H_i$ is the in-plane Hamiltonian of the i$^{th}$ graphene layer (the first, second and third graphene layers refer to the top, middle and bottom layers in TLG, respectively.); $k$ and $\sigma$ denote the momentum vector and spin of electrons in graphene, respectively; $a^+$ ($b^+$) and $a$ ($b$) are the creation and annihilation operators of electron at A (B) sublattice, respectively; $\vec{a}_1$ and $\vec{a}_2$ are the basis lattice vectors of graphene; $H_{12}$ and $H_{23}$ are the Hamiltonians due to interlayer coupling; $t$ is the in-plane nearest-neighbor hopping energy. We note that the coupling between the top two layers is the same for TLG with different stacking order, as the top two layers are AB-stacked. Thus, $H_{12}$ can be written as following:

$$H_{12} = -\gamma \sum_{k,\sigma} \left( a^+_{1,k,\sigma} b_{2,k,\sigma} + h.c. \right) \quad (4)$$

where $\gamma$ is the interlayer nearest-neighbor hopping energy, which is about one tenth of the in-plane nearest-neighbor hopping energy $t$.[5] Meanwhile, the coupling between the bottom two layers in TLG depends on their stacking order. Thus, for different type of TLG the interlayer Hamiltonian $H_{23}$ in Eq. (1) can be written as

$$H_{23}^{ABA} = -\gamma \sum_{k,\sigma} \left( b^+_{2,k,\sigma} a_{3,k,\sigma} + h.c. \right) \quad (5)$$

$$H_{23}^{ABC} = -\gamma \sum_{k,\sigma} \left( a^+_{2,k,\sigma} b_{3,k,\sigma} + h.c. \right) \quad (6)$$

$$H_{23}^{ABB} = -\gamma \sum_{k,\sigma} \left( b_{2,k,\sigma}^+ b_{3,k,\sigma} + h.c. \right) \quad (7)$$

The band structure and DOS of TLG with different stacking order can be calculated from these Hamiltonians.

Figure 4(a) shows the band structure of ABA-stacked TLG around the Dirac point K. Two linear bands (red curves) and two parabolic bands (green curves) intersect at the Dirac point. This band structure leads to a V-shaped DOS (Fig. 4(d)) with a minimum at $E_F$, in agreement with the d$I$/d$V$ spectrum of TLG on Ru shown in Fig. 3(a). For the ABC-stacked TLG, the two parabolic bands (green curves) with lowest energies intersect at the Dirac point. This leads to the flat bands around $E_F$ (Figs. 4(b)) and thus a pronounced peak at $E_F$ in the corresponding DOS (Figs. 4(e)), in line with the d$I$/d$V$ spectrum shown in Fig. 3(b). Fig. 4(c) and (f) show the calculated band structure and DOS of ABB-stacked TLG, respectively. The two bands (green curves) with lowest energies intersect three times around the Dirac point, resulting in two sharp peaks at ±68 meV with a minimum at $E_F$ in the corresponding DOS. The calculated electronic structure is consistent with the d$I$/d$V$ spectrum displayed in Fig. 3(c), except the difference in the interval between the two sharp peaks. We note that the strain in the bottom layer induced by the strong coupling between the graphene layer and the Ru(0001) substrate was not taken into account in our TBA calculations, which might count for the difference between the calculated and measured DOS of TLG with ABB-stacking. The singularities at about 0.3~0.4 eV might be due to the simplified model that we have used. Thus, the d$I$/d$V$ spectra shown in Fig. 3(a), (b) and (c) are assigned to TLG with ABA-, ABC- and ABB-stacking, respectively. Recently, Lalmi *et al.* prepared flow-shaped TLG on SiC with ABC-stacking, and observed a pronounced peak around $E_F$ in the d$I$/d$V$ spectra by STM/STS measurements.[12] Xu and coworkers obtained ABC-stacked TLG on HOPG surface and collected similar sharp peak in vicinity to $E_F$ in the d$I$/d$V$ spectra.[34] Our results are in line with these reports, evidencing the flat bands close to $E_F$ in ABC-stacked TLG. It is noteworthy that the ABB-stacking is energetically unfavorable for free-standing TLG. The ABB-stacked TLG in our work is stabilized by the Ru(0001) substrate. The flat bands with fine structure of ABB-stacked TLG may give rise to novel electron correlation phenomena.

In summary, we have studied the structural and electronic properties of TLG epitaxially grown on Ru(0001) surface by LT-STM/STS combined with TBA calculations. TLG on Ru(0001) shows a flat surface due to the screening effect of the bottom two layers. The AB-stacking in the top two layers and the coexistence of AA- and AB-stacking in the bottom two layers lead to three different stacking orders of TLG, namely, ABA-, ABC- and ABB-stacking. STS measurements combined with TBA calculations reveal that the DOS of TLG with ABC- and ABB-stacking are characterized by one and two sharp peaks near to the Fermi level, respectively, in contrast to the V-shaped feature of TLG with ABA-stacking. Our work demonstrates that TLG on Ru(0001) might be an ideal platform for exploring stacking-dependent electronic properties of graphene.

The work was financially supported by grants from MOST (Nos. 2015CB921103), NSFC (Nos. 61574170) and CAS in China.

**Figure captions**

**Figure 1**. STM images of TLG grown on Ru(0001). (a) Over view showing the flat TLG islands surrounded by corrugated BLG (sample bias: $U$ = -900 mV; tunneling current: $I$ = 12 pA). (b) Atomic-resolution image showing the coexistence of honeycomb and hexagonal lattices of the atop regions of BLG ($U$ = -110 mV, $I$ = 11 pA). (c) Atomic-resolution image showing the hexagonal lattice of TLG ($U$ = -190 mV, $I$ = 20 pA). (d) Atomic-resolution image showing the seamless connection between the top layer of BLG and TLG ($U$ = -20 mV, $I$ = 52 pA). (e) Schematic model of TLG on the lower terrace and BLG on the upper terrace of an atomic step of Ru(0001) surface.

**Figure 2**. Structural model of TLG on Ru(0001). The close-ups show three types of stacking order, namely, ABA-, ABC- and ABB-stacking. For simplicity, the Ru(0001) substrate is not shown. The $<10\bar{1}0>$ directions of all layers are parallel, whereas the lattice of the bottom layer is stretched with respective to the two top layers. For clarity, the strain of tension in the bottom layer is enlarged to reduce the size of the model.

**Figure 3.** Three types of representative d$I$/d$V$ spectra collected from different positions of TLG on Ru(0001) surface ($U$ = -400 mV, $I$ = 100 pA).

**Figure 4**. Calculated band structures around K point (a-c) and DOS (d-f) of free-standing TLG with ABA-, ABC- and ABB-stacking, respectively. In all TBA calculations, only the nearest

neighbor hopping was taken into account. The in-plane and out-of-plane hoping parameters are $t = 3.16$ eV, and $\gamma = 0.32$ eV, respectively.

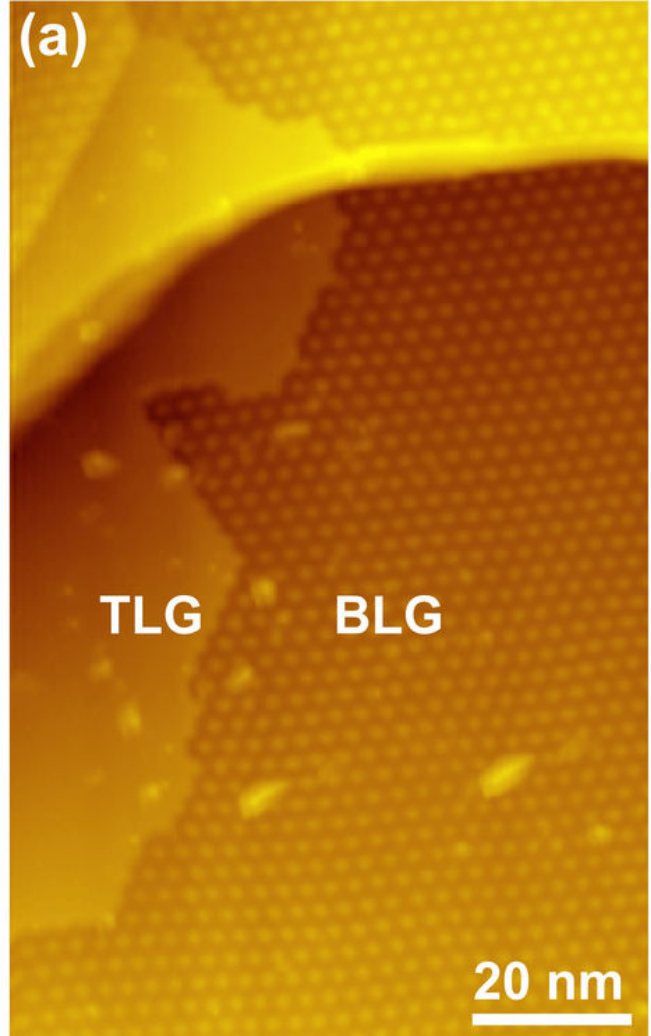
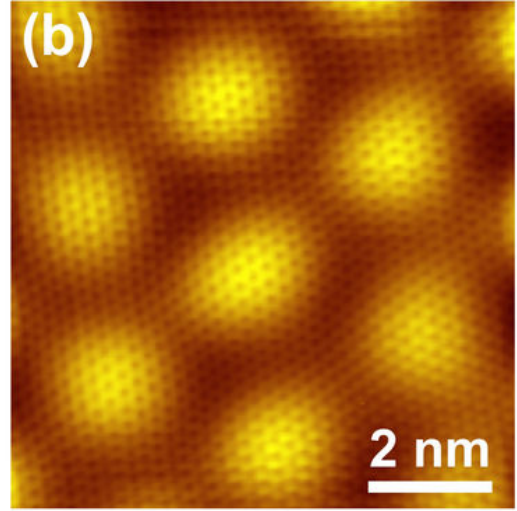
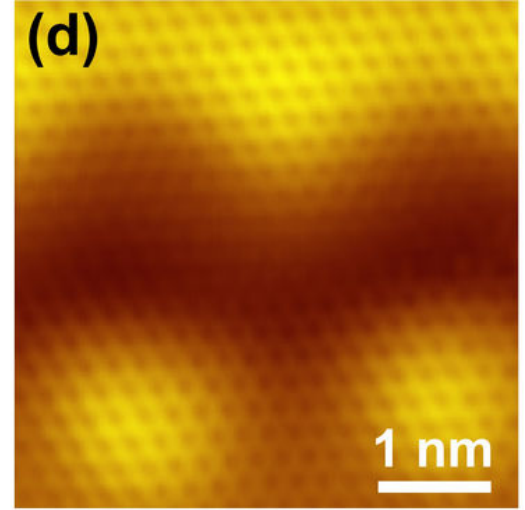
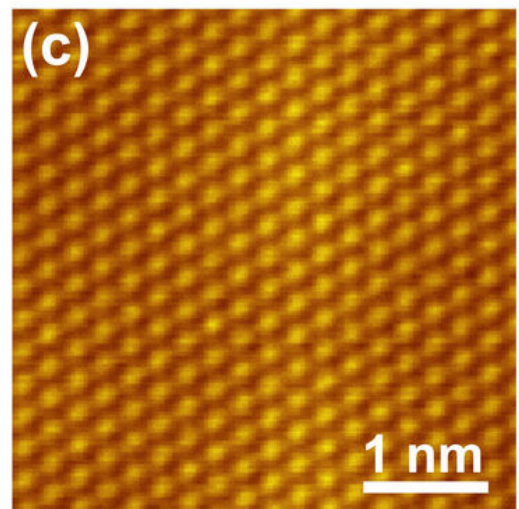
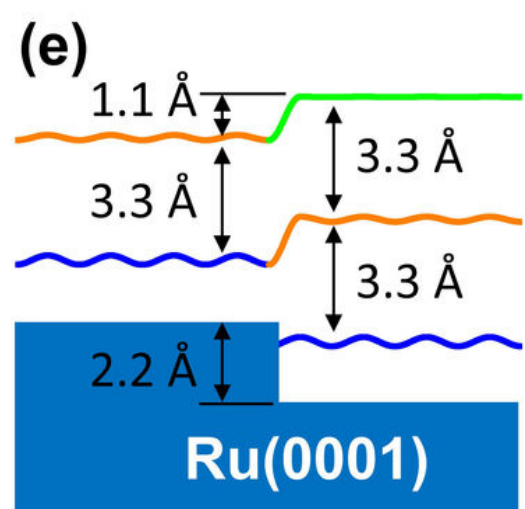

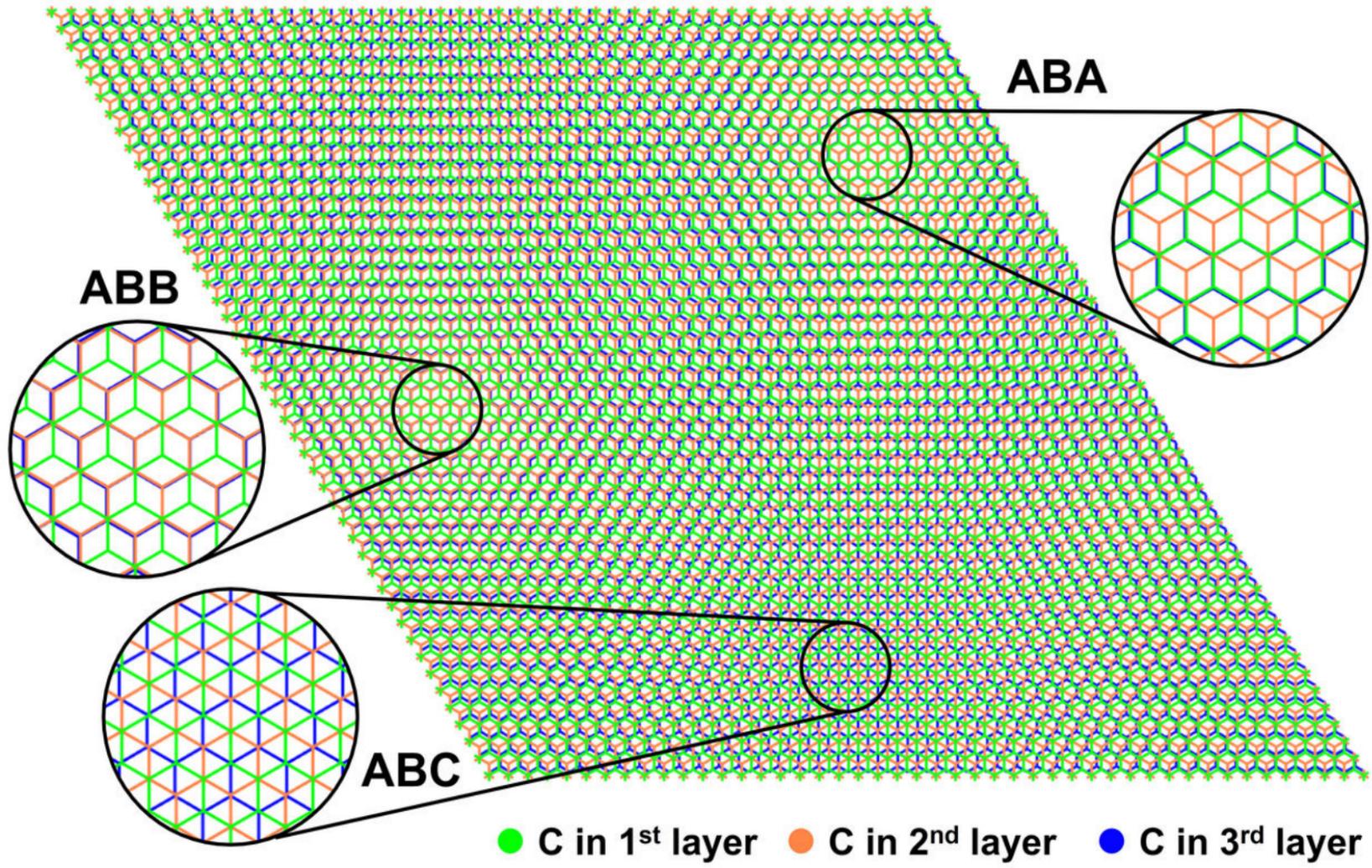

● C in 1st layer   ● C in 2nd layer   ● C in 3rd layer

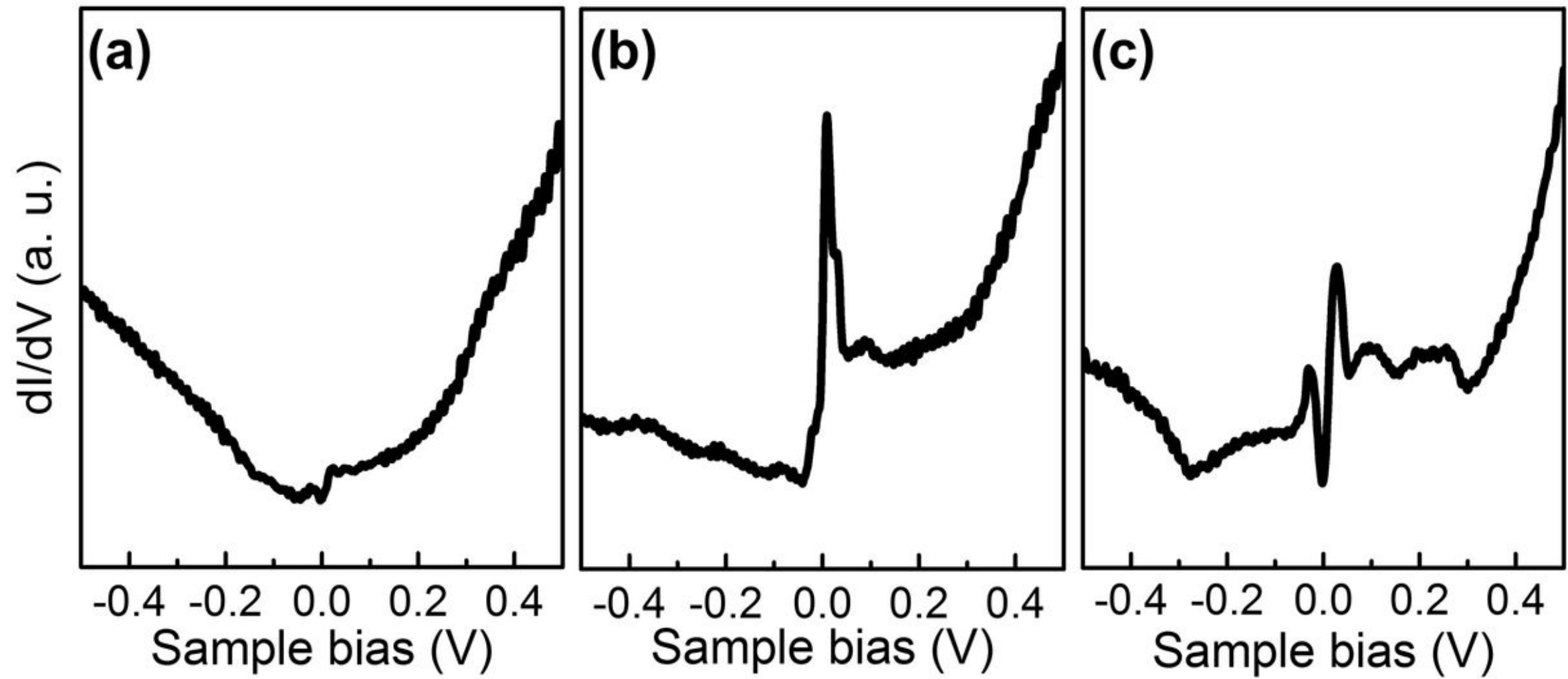

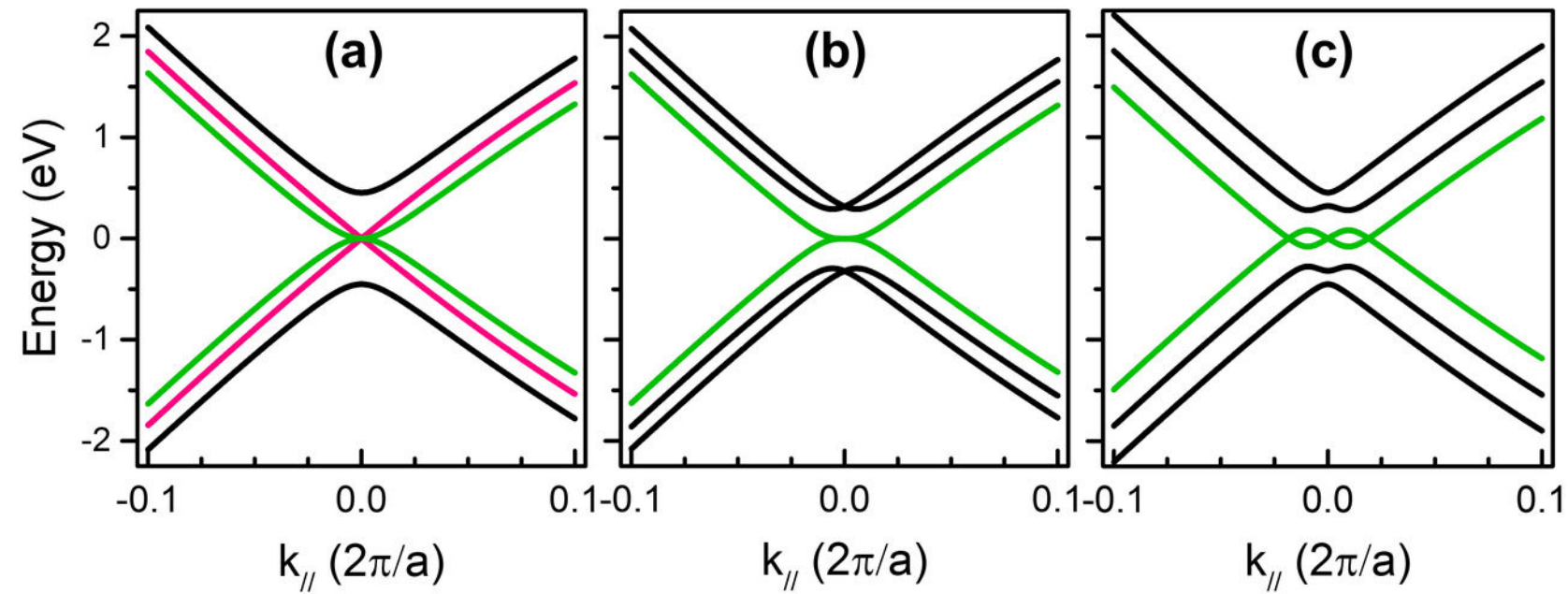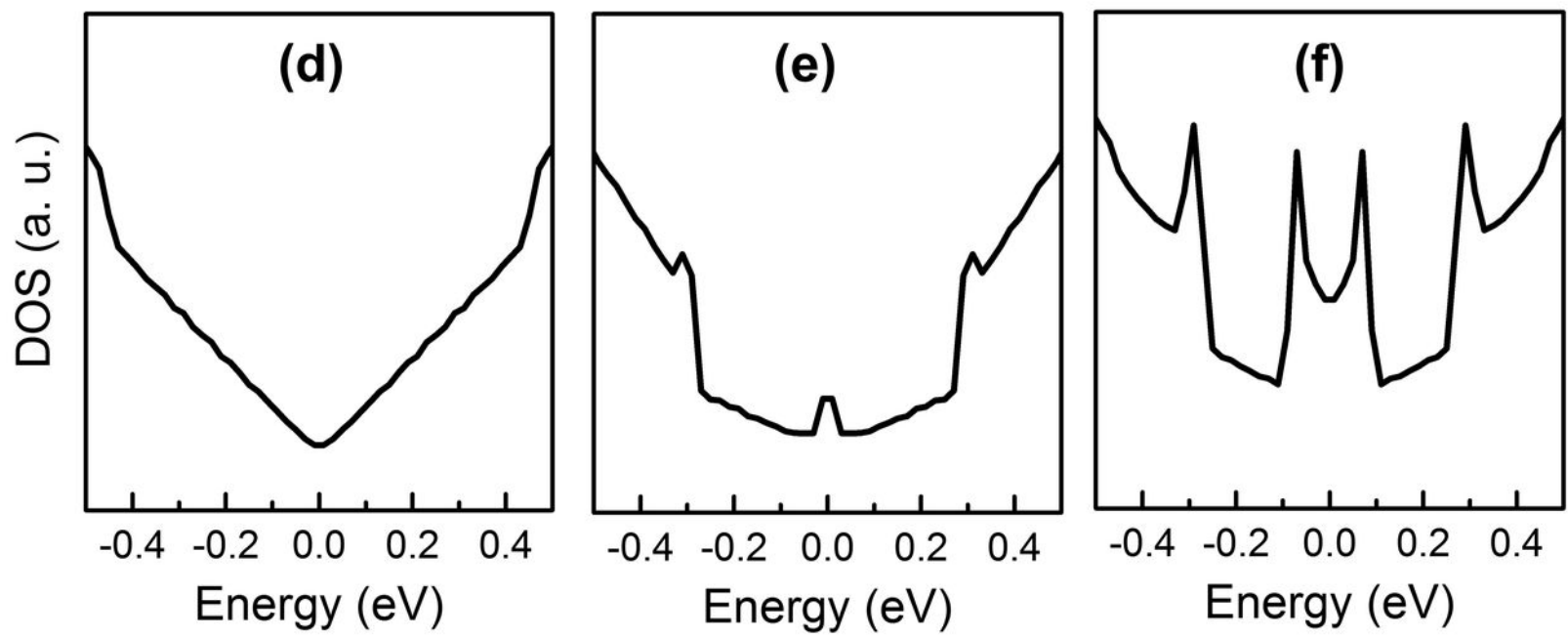